\documentclass{aastex}
\usepackage{emulateapj5,apjfonts}
 
\begin{document}

\def\pdot {\dot P}
\def\Omdot {\dot \Omega}
\def\ltsima{$\; \buildrel < \over \sim \;$}
\def\lsim{\lower.5ex\hbox{\ltsima}}
\def\gtsima{$\; \buildrel > \over \sim \;$}
\def\gsim{\lower.5ex\hbox{\gtsima}}
\def\msole{~M_{\odot}}
\def\mdot {\dot M}
\def\ee {1E~1207--5209~}
\def\cha {\textit{Chandra~}}
\def\gg {G 296.5+10.0~}
\def\xmm  {\textit{XMM-Newton~}}

\title{Pulse phase variations of the X--ray spectral features in the
radio-quiet neutron star \ee\ }
\author{S. Mereghetti, A. De Luca, P.A. Caraveo}
\affil{Istituto di Astrofisica Spaziale e Fisica Cosmica, Sezione di Milano  ''G.Occhialini'' - CNR \\
v.Bassini 15, I-20133 Milano, Italy \\
sandro@mi.iasf.cnr.it}
\author{W. Becker }
\affil{  Max Planck Institut f\"{u}r Extraterrestrische Physik, Giessenbachstrasse,\\
Postfach 1312, D-85740, Garching, Germany \\}
\author{R. Mignani}
\affil{ European Southern Observatory, Karl Schwarzschild Strasse 2,  \\
 D-85740, Garching, Germany \\}
\author{G.F. Bignami}
\affil{Agenzia Spaziale Italiana, v. Liegi 26, Roma, I-00198, Italy \\
and Universit$\grave{a}$ degli Studi di Pavia  }

\begin{abstract}

We present the    results of an \xmm  observation of the radio-quiet
X--ray pulsar \ee\ located at the center of the shell-like
supernova remnant  \gg\ . The X--ray spectrum is characterized by the presence of two
phase-dependent absorption lines at energies $\sim$0.7 keV and $\sim$1.4 keV.
Moreover, these broad spectral features have significant substructure,
suggesting that they are due to the blending of several  narrower lines.
We interpret such features as evidence for an atmosphere containing metals
and a magnetic field value of a few 10$^{12}$ G,
consistent with the observed spin-down rate
$\pdot$=(1.98$\pm$0.83)$\times$10$^{-14}$ s s$^{-1}$.
Since \ee\ is the only X--ray emitting pulsar showing evidence of such features,
we tentatively link them to the unique combination of  age and
energetics that characterize this object.
We suggest that a young age and a low level of magnetospheric
activity are  favorable conditions for the detection of atomic spectral features
from $Z>1$ elements in neutron star atmospheres, which would be either blanketed by a thin layer of
accreted hydrogen in older objects or masked by non-thermal processes in young
energetic pulsars.

\end{abstract}
 

\section{Introduction}

The thermal emission from the surface of a neutron star
traces  the star's cooling history. Its study
can thus provide invaluable information on the poorly known
equation of state of matter at super-nuclear densities and on
physical processes in strong magnetic fields. Satellite
observations carried out in the last decade have clearly
shown the thermal origin of the soft X--ray emission
($\sim$0.1-3 keV) from a handful of middle-aged radio pulsars
(Becker \& Tr\"{u}mper 1997) as well as from several
radio-quiet neutron stars (Caraveo, Bignami \& Tr\"{u}mper 1996;
Treves et al. 2000).

It is expected   that the presence of an
atmosphere on the   neutron star surface
distort the emerging radiation  by altering the
blackbody energy distribution and  introducing absorption  features
(see, e.g., Zavlin \&  Pavlov 2002 for a recent review).
Fits with atmospheric models generally yield
emitting regions compatible with standard neutron star dimensions,
thus providing indirect evidence for the presence of an atmosphere.
However, until recently no convincing evidence for the absorption features
predicted by these models was found, thus leaving
substantial uncertainty on the atmospheric composition and magnetic field.

Here we report on an \xmm observation of the radio-quiet neutron star  \ee .
Previous X--ray, optical and radio observations
(Bignami, Caraveo \& Mereghetti 1992; Mereghetti, Bignami \& Caraveo 1996;
Vasisht et al. 1997)
strongly suggested a  neutron star nature for  this source, located close to
the geometrical center of the shell-like
supernova remnant  \gg  (Roger et al. 1988).
This was confirmed by the discovery  of fast  X--ray  pulsations with
period $P$=0.424 s (Zavlin et al. 2000).
A  subsequent    measurement (at the $\lsim2\sigma$ level)
of a positive  period derivative
$\pdot$ = (2.0$^{+1.1}_{-1.3}$) $\times$ 10$^{-14}$ s s$^{-1}$  (Pavlov et al. 2002a)
results in a large discrepancy between the pulsar's
characteristic age $\tau_{c}\equiv\frac{P}{2\pdot}=200-900$ kyrs
and  the age  of  7 kyrs (with a factor 3 uncertainty) estimated for the
associated SNR (Roger et al. 1988).

Our  \xmm data show the presence  of broad absorption features at
$\sim$0.7 and $\sim$1.4 keV in the spectrum of  \ee\ .
Such features have been independently discovered in data from the
\cha satellite (Sanwal et al. 2002).
The high throughput of the \xmm telescope, coupled to the good spectral and timing
resolution of the
European Photon Imaging Camera (EPIC) instrument,
allow us to show  that the lines have a significant substructure, which varies with the
phase of the pulsar.

\section{Data Analysis and Results}

The \xmm\ observation of \ee\ started on December 23, 2001 at 19:13 UT
and lasted 28.4  ks.
We concentrate here on data obtained with the
EPIC instrument, which consists of two MOS CCD detectors (Turner et al. 2001) and a PN
CCD instrument (Str\"{u}der et al. 2001), for a total collecting
area $\gtrsim$ 2500 cm$^2$ at 1.5 keV.

The PN camera was operated in Small Window mode in order to have  a time resolution
(6 ms) adequate to study the pulsations without sacrificing the imaging,
while  both  MOS  CCDs were in the Full Frame mode (2.6 s resolution).
All the detectors used the Medium thickness filter.
All the data were processed with the \xmm\ Science Analysis Software (SAS Version 5.3).
After screening  to remove time intervals  of high proton background  which
affected the MOS data, and correcting for the dead time, we obtained
net exposure times of 18.7 ks, 22.3 ks and 24.8 ks in the PN,  MOS1 and MOS2 respectively.

\ee\ was clearly detected at a position consistent with previous
measurements (Mereghetti et al. 1996), with net count rates (0.3--3 keV)
of   1.352$\pm$0.008 counts s$^{-1}$
and  0.362$\pm$0.004 counts s$^{-1}$, respectively in the PN and in each MOS.

A circular extraction region with radius 40$''$ was used for the
timing and spectral analysis of the PN data.
The background shows a slight intensity gradient across the chip.
We verified that using background regions at different positions
did not affect significantly
the best fit parameters (the source  count rate is $\sim$30 times greater than that of the
background in the 0.3--3 keV   range). We therefore finally used
for the background spectral extraction  a  box of dimensions
4$'\times2'$ located to the north of the source.

For both MOS cameras we used a circular source extraction region (radius 40$''$)
and the background spectrum was estimated from a concentric annulus with radii 75$''$
and 200$''$. This region is entirely contained in the central chip and is not
affected by  contamination from the  internal Si-K fluorescence, largely present
near the CCD edges.

\subsection{Timing Analysis}

For the timing analysis we used only the PN counts with pattern 0-4 and with energy in the
range 0.2-2.5 keV.
The times of arrival  were converted to the Solar System Barycenter
and folded, with 8 phase bins, in a range of trial periods around the expected value.
This gave a significant detection of the pulsation with a maximum $\chi^2$=110
at P=424.131 ms.
To determine more accurately the period value we
fitted the $\chi^{2}$ versus trial period curve with the appropriate $sinc^{2}$ function,
and computed the  error on P using the relation between maximum $\chi^{2}$ and uncertainty
derived by Leahy (1987). This resulted in our best estimate of
P = 424.13084$\pm$0.00046 ms.
For an independent assessment of the period uncertainty, we generated
artificial  data sets with the same properties (duration, number of counts,
pulsed fraction, etc...) of our observation and analyzed
them in the same way. The difference
between the derived and the true period was found to have a Gaussian distribution
with $\sigma=5\times10^{-7}$ s, thus confirming the above
estimate of the period uncertainty.

Comparison with the period   measured  in January  2000 with \cha
(as reported in the re-analysis of Pavlov et al. 2002a)
yields a period derivative  
$\pdot$=(1.98$\pm$0.83)$\times$10$^{-14}$ s s$^{-1}$.

The folded light curve is nearly sinusoidal, with a $\sim$8\% pulsed fraction (Fig. 1)
and no significant energy dependence.
We have also examined the  profiles in the 0.3--1 keV and 1--1.7 keV energy bands
(16336 and 7639 counts, respectively)
without finding the  phase shift reported by Pavlov et al. (2002a). 
Simulations show that a phase shift of $\sim$0.4--0.5 has a probability 
much smaller than 1\% to be obtained by chance with the available  statistics.
This suggest that the pulsar light profile might be time variable.

\subsection{Spectral Analysis}

All spectral modeling was done with XSPEC V11 and using the most recent EPIC
response matrices \footnote{PN version 6.1 epn\_sw20\_sdY9\_medium, epn\_sw20\_sY9\_medium;
MOS m1\_med\_v9q20t5r6\_all\_15.rsp, m2\_med\_v9q20t5r6\_all\_15.rsp}.
For the PN  analysis we used  both single and double events
in the 0.3--3 keV energy range
(we checked \textit{a posteriori} that the results do not change
by using only single events).
The spectra were binned  to have at least 40 counts per channel and to
over-sample by a factor 3 the instrumental energy resolution.

As found by Sanwal et al. (2002),
fits with single-component models (power law, thermal bremsstrahlung, blackbody)
gave unacceptable results ($\chi^{2}$/dof $>$ 4.4, dof=81)
due to the presence of broad absorption features at $\sim$0.7 and $\sim$1.4 keV.
For completeness, we also tried the atmospheric models  available in the
XSPEC spectral fitting package, although they
refer to weakly  magnetized   neutron stars (B$\lsim10^{9}-10^{10}$ G)
and might not be appropriate in the case of \ee , where the
pulsations and the measured $\pdot$ testify the presence of a higher
magnetic field (B$\sim3\times10^{12}$ G).
Hydrogen-atmosphere models
(Zavlin, Pavlov \& Shibanov 1996),
which do not predict lines in the observed energy range,
gave residuals very similar to those of the blackbody model, but, as expected,
a lower temperature (kT$\sim$0.09 keV, as measured by an observer at infinity)
and a larger emitting region.
Models implying
atmospheres with solar abundance or iron composition  (G\"{a}nsicke, Braje \& Romani 2002)
gave unacceptable fits: the absorption lines they
predict are too narrow to reproduce the features visible in the EPIC spectra.

%
\centerline{\includegraphics[width=7.5cm,angle=-90]{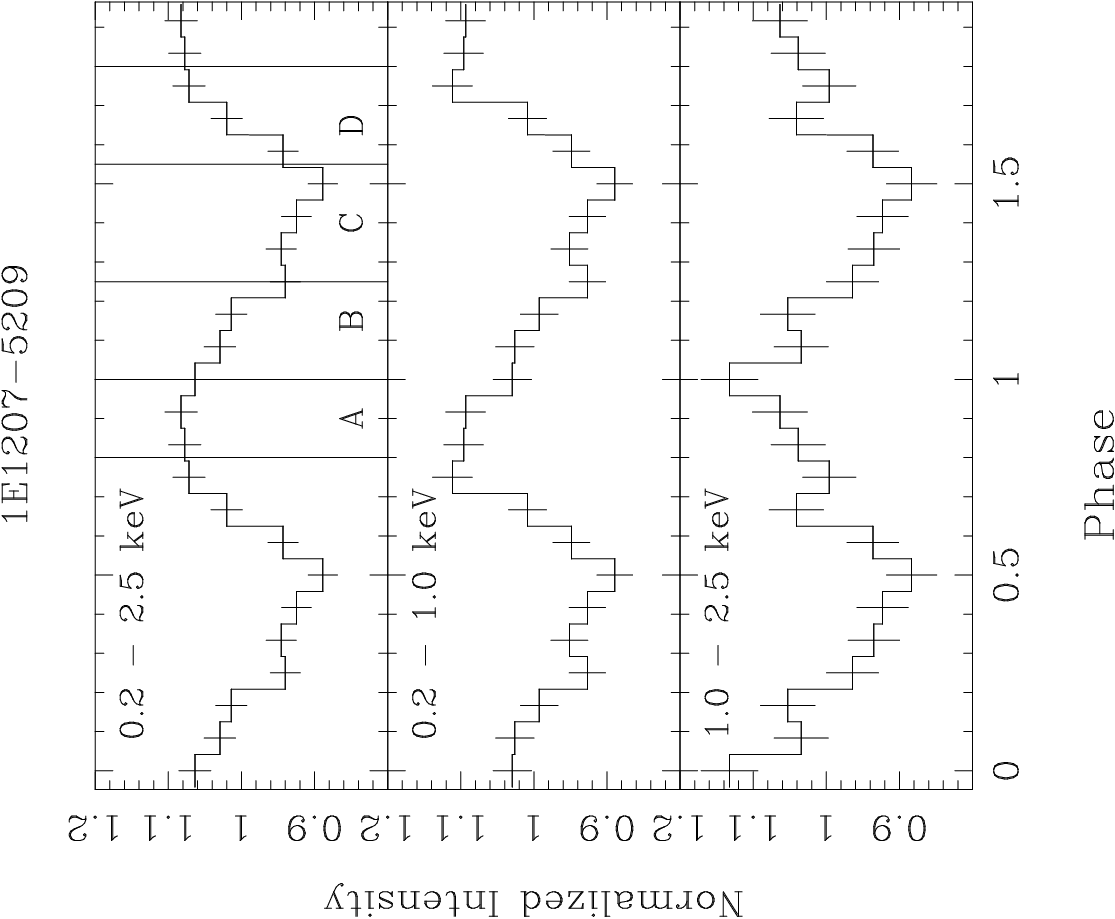}}
\figcaption{Light curve of \ee for three different  energy ranges.
}
\label{}
\centerline{}

Acceptable fits could only be obtained by adding to the models two absorption lines,
which for simplicity we modeled with Gaussian profiles.
As shown in Table 1, the line parameters are only slightly dependent on the
model adopted for the continuum.
The single blackbody   yields a lower interstellar absorption, while the other models
give higher values, closer to the estimates obtained from
radio observations (Giacani et al. 2000).
Although  the
blackbody plus power-law gives formally the best fit (see Fig. 2),
we believe that this simply reflects the shortcomings of other models alone
to reproduce the low energy part of this complex atmospheric spectrum, rather than
being evidence for a distinct non-thermal component.

The fit residuals  (lowest panel of Fig. 2) show that the broad lines
have a profile characterized by the presence of significant sub-structure.
Deviations from a smooth gaussian profile are seen at $\sim$0.6 keV, $\sim$0.75 keV
and $\sim$0.85 keV. Although the exact
significance of such features is difficult to quantify, this may imply that
the broad absorption below 1 keV   is produced by  blending of several narrower lines.

Another interesting feature is shown by the residuals near 2 keV. This was also noticed
in the \cha spectrum by Sanwal et al. (2002), who could not exclude an instrumental
effect. The fact that the same line is possibly present   in the \xmm data,
even if at low significance, indicates that it might be really present in the source.

We performed a similar spectral analysis based on the data from the two MOS cameras.
This led to results entirely consistent with the ones discussed above, but with
larger uncertainties caused by the lower counting rate. All the strongest
spectral features were clearly visible in both CCDs.

\centerline{}
\centerline{\includegraphics[width=6.5cm]{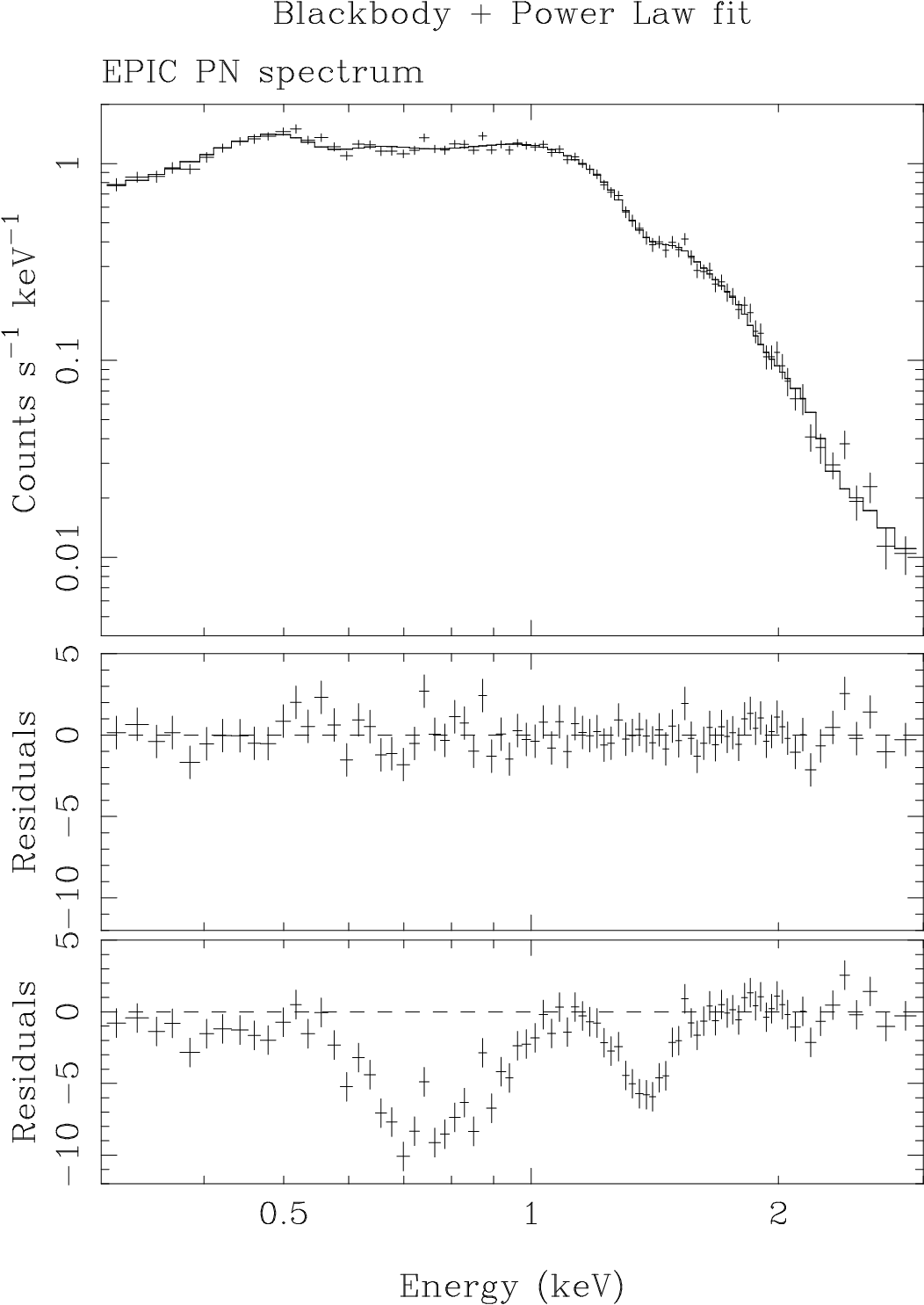}}
\figcaption{
Fit    with a    blackbody plus
power law and two gaussians.
In the upper panel the data are compared to the model folded through
the instrumental response.  The middle panel shows the residuals in units of sigma, 
the lowest one shows the residuals obtained by removing the lines from the model. }
\label{}
%

\subsection{Phase-resolved spectroscopy}

To search for possible phase-dependent spectral variations, we divided the PN
data in four sets corresponding to the phase intervals   indicated
by the vertical lines in Fig. 1.
The resulting spectra
were fitted with  blackbody models,    keeping the absorption fixed at the
best fit value of the average spectrum (N$_{H}$=3$\times$10$^{20}$ cm$^{-2}$).
While similar temperatures were obtained in the four phase intervals,
the fit residuals (Fig.~3) clearly indicate that  the absorption
features are phase-dependent.
In particular the line at $\sim$1.4 keV
is virtually  absent at the pulse peak and is more pronounced during
the minimum and the rising parts of the pulse profile (see Table 2).
Shape variations are also visible  for the
lower energy feature, which shows a variable sub-structure,
suggesting that several narrow lines
contribute differently at the various phases.

\centerline{}
\centerline{\includegraphics[width=6.5cm,angle=-90]{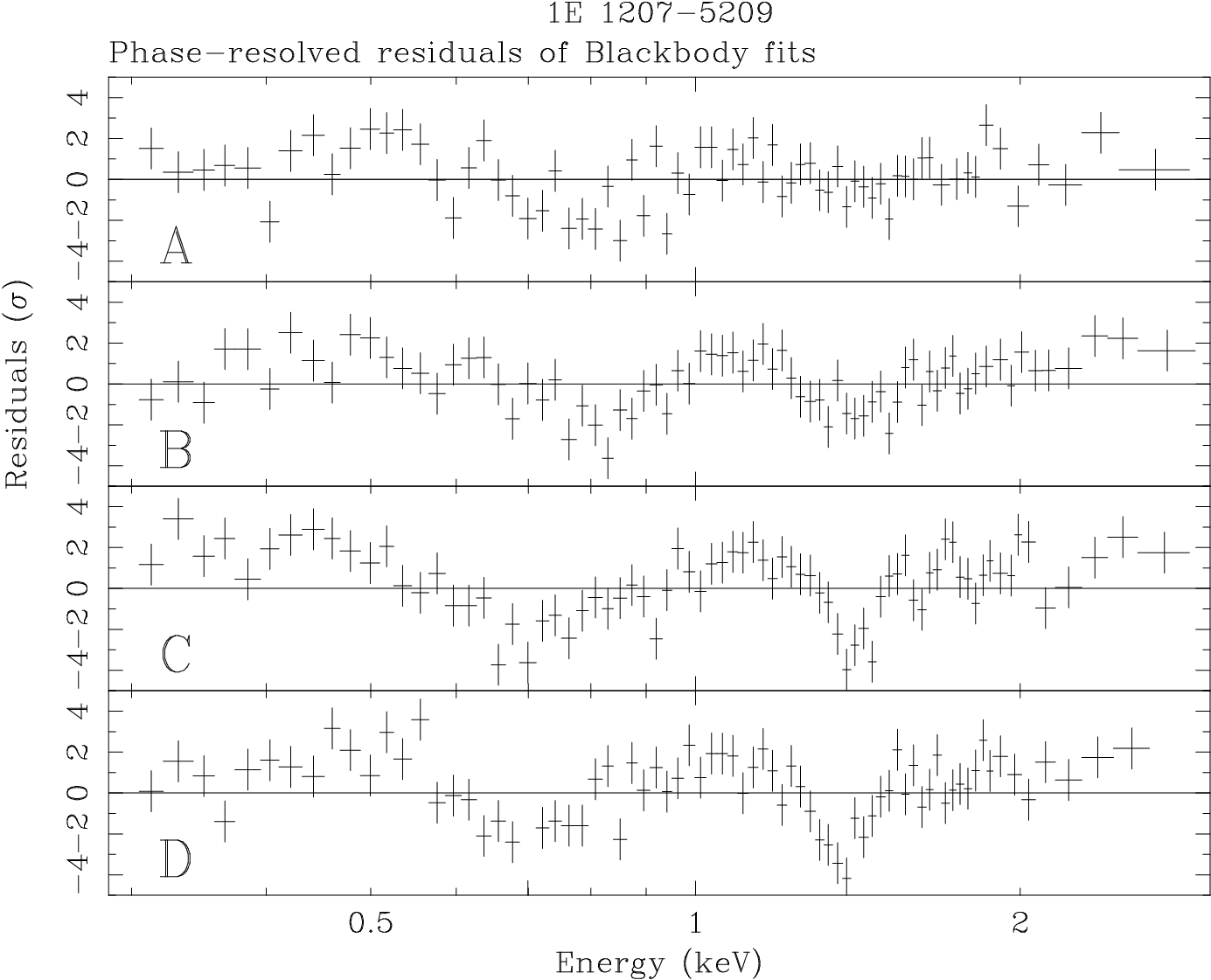}}
\figcaption{
The four panels show the residuals of
blackbody fits with N$_{H}$ fixed at 3$\times$10$^{20}$ cm$^{-2}$
for the phase intervals shown in Fig. 1.
}
\label{}
\centerline{}

\section{Discussion}

Our \xmm observation confirms the presence of spectral features in the spectrum of \ee\ ,
as recently reported by Sanwal et al. (2002), and allows us to study them in more
detail. In particular, both the PN and the two MOS spectra indicate that  the  feature
at $\sim$0.7 keV cannot be well described by a single gaussian line due to the presence
of significant substructure. The  PN spectra show phase-dependent variations,
particularly pronounced for what concerns the  intensity and width of the line at $\sim$1.4 keV.
The presence of phase-dependent variations, leads to the obvious, but important
conclusion that the lines are linked to the rotating neutron star and not due
to absorption in the line of sight.
 
According to Sanwal et al., an interpretation of  these spectral  features
in terms of cyclotron resonance lines is difficult, considering the
magnetic field inferred from the $\pdot$ measurement, B$\sim$3$\times10^{12}$ G,
and the relative intensity of the two lines.
This view has been criticized by Xu et al. (2002), who considered the possibility
that \ee\ be a bare strange star, spinning down in the propeller regime due to
the presence of a fossil disk.

Considering the conventional scenario of an isolated neutron star,
a more likely possibility, also suggested by the structured and phase-dependent
profiles seen with EPIC,  is that these   features result from atomic transitions.
Pavlov et al. (2002b) interpret them as HeII lines in a strong
magnetic field (B$\sim$(1.4--1.7)$\times$10$^{14}$ G), and derive for the neutron star
a radius  to mass ratio of (8.8--14.2) km $\msole^{-1}$.
However, such a  magnetic field is much higher than  the value
(3$\pm$0.6)$\times$10$^{12}$ G estimated, assuming magnetic dipole braking,
from the spin-down value (now confirmed by the \xmm data).
This implies  either that the
observed $\pdot$ is affected by
glitches and/or significant timing noise
(and the true spin-down is of the order of $\pdot$$\sim6\times10^{-11}$ s s$^{-1}$)
or that  field components stronger than the dipole   are present
on (some region of) the neutron star surface (e.g. due to an off-centered dipole).

Alternatively, the lines can be due to heavier elements in a more conventional
magnetic field.
For example, models of  magnetized (B$\gsim10^{12}$ G) iron atmospheres
(Rajagopal, Romani \&  Miller 1997)
predict, for temperatures $\sim10^{6}$ K, many
absorption lines due to atomic transitions in the range above 0.3 keV.
Owing to the
magnetic field and temperature variation across the neutron star surface,
these lines will probably be blurred.
Indeed, the results of our phase-resolved spectroscopy show that different
physical conditions, most likely due to changing magnetic field configurations,
are present on the neutron star regions responsible for the
emission visible at different  phases.
As a consequence,  the physical parameters inferred from the
fit to   phase-averaged
spectra should be taken with some caution.
It is also likely that the presence of several atomic absorption lines and
edges affect the
the parameters of the continuum as derived from medium resolution spectra.
\footnote{After this paper was submitted, Hailey \& Mori (2002) proposed that the features
in \ee are produced by He-like Oxigen or Neon and predicted the presence of substructures in the lines.}

\ee\ is the only neutron star in which significant X--ray absorption
features have been detected so far (excluding of course the bright
neutron stars accreting in binary systems).
Observations with high statistics and good spectral resolution have been
recently carried out for several thermally emitting neutron stars of various classes:
the middle aged radio pulsar PSR B0656+14 has a spectrum
well described by two blackbody components,
without spectral lines in the 0.15--0.8 keV range (Marshall \& Schultz 2002).
No lines  were found in the 8.4 s radio-quiet
pulsar RX J0720.4--3125 (Paerels et al. 2001) and in the nearby
isolated neutron star  RX J1856.5--3754. The latter was observed
for more than 500 ks with \cha and
its spectrum was found to be
well described by a blackbody function with temperature kT$_{BB}$=61 eV
without evidence for any of the spectral lines or edges predicted
by the models (Drake et al. 2002). No lines have been detected   in
the  spectra of the Vela pulsar (Pavlov et al. 2001a)
and  of the millisecond pulsar PSR J0437--4715 (Zavlin et al. 2002),
which are comparable in  quality  and statistics
to the data presented here.
Why is \ee\ different from the other
cooling neutron stars which have been deeply scrutinized for the presence of
lines with negative results?

Before trying to answer this question, we must address the problem of the age
of \ee. In this respect, its association with \gg\ is crucial.
This SNR has a remarkable bilateral symmetry. The two radio arcs that compose the
shell have different curvatures, indicating a larger expansion velocity
for the ejecta on the western side. The likely site of the supernova explosion
is thus relatively well determined to lie to the east of the
geometric  center, within 8$'$ from the position
of the pulsar.  The spatial coincidence between the pulsar and the SNR
is remarkable, also in view of the
relatively small number of such objects at this galactic latitude (b=10$^{\circ}$).
Furthermore, HI observations of this region (Giacani et al. 2000) indicate a spatial
correlation between the two objects also for what concerns their distance.
Thus we can consider this as one of the strongest neutron star/SNR association.
An age of 7 kyrs has been derived for  \gg\  (Roger et al. 1988),
very different from the pulsar characteristic age
$\tau_{c}=(340\pm140)$ kyrs.
Although the SNR age estimate depends on several uncertain parameters,
like its linear size,
initial kinetic energy ($E_{51}\times10^{51}$ ergs)
and the density of the interstellar medium ($n$ cm$^{-3}$),
it is very unlikely that \gg\ is older than a few $\times10^{4}$ years.
For example, even taking the maximum distance of 4 kpc compatible with the HI
observations (Giacani et al. 2000), an age of $\gsim$100 kyrs can be obtained only for
$(n / E_{51}) \gsim$1.
This would require a very small initial kinetic energy, since at this distance the SNR
would be $\sim$700 pc above the
galactic plane, where the interstellar density is
expected to be very small.
In conclusion, we believe that a distance of $\sim2$ kpc and an age of the order of
$\sim10^{4}$ years are more likely values.

Although we cannot rule out other possibilities (e.g. one or more glitches in the
last two years, significant timing noise, or a faint binary companion),
the simplest explanation to reconcile $\tau_{c}$ with a young age is that \ee\ was
born spinning at a period close to its current value.

Thus  \ee\ appears as a unique example of a young pulsar with a small
rotational energy loss.
Its rotational energy loss, $\dot{E}_{rot}$$\sim10^{34}$ erg s$^{-1}$,
is the smallest one of all the radio/X--ray  pulsars
with reliable SNR's associations (Kaspi \& Helfand 2002).
Such  ''Crab-like'' or ''Vela-like'' pulsars  have typically $\dot{E}_{rot}>10^{36}$
erg s$^{-1}$ and X--ray luminosity in the range 10$^{32}$-10$^{37}$  erg s$^{-1}$
(see, e.g., Possenti et al. 2002).
Indeed, no plerion has been detected either in radio or X--rays,
nor any hint has been found of  $\gamma$-ray emission from
our object.
On this ground, the comparison with the similar age Vela pulsar is striking:
beside being active as a radio pulsar, Vela is
also the brightest source in the sky at E$>$100 MeV,
and is powering a bright radio/X--ray synchrotron nebula.
If the   lines   are the signature of the presence of metals in the atmosphere,
their absence in the older objects mentioned above
can be explained by a small layer of  hydrogen accreted, even at a modest
rate from the interstellar medium over a time span of a few hundred kyrs.
The absence of lines (or the difficulty to detect them) in young but
energetic pulsars like Vela could be due to the possible perturbing effects
of the non-thermal particles accelerated in  the magnetosphere.

\section{Conclusions}

The \xmm observations reported here show unambiguously that the
absorption  features independently discovered with \cha (Sanwal et al. 2002)
in the X--ray spectrum of \ee are phase-dependent and
have significant sub-structure.
This supports an
interpretation in terms of atomic transitions in regions with different temperature
and magnetic field on the neutron star surface.
A detailed analysis of phase-resolved spectra with higher statistical quality
will undoubtedly provide important information on the
surface composition and other  neutron star parameters.

The  period value reported here is  identical, within the uncertainties, to that obtained
two weeks later (Pavlov et al. 2002a), but our smaller error reduces the uncertainty
on the pulsar spin-down rate
$\pdot$=(1.98$\pm$0.83)$\times$10$^{-14}$ s s$^{-1}$.
Since the association with the relatively young SNR remnant is extremely likely, we
consider the high characteristic age of \ee\ as evidence that this pulsar
was born with a long period.

We suggest that the combination of a young age and a low level of magnetospheric
activity are  favorable conditions for the detection of atomic spectral features
from Z$>$1 elements, 
which would be either blanketed by a thin layer of
accreted hydrogen in older objects or masked by non-thermal processes in young
energetic pulsars.
If this scenario is correct,  the most  promising candidates for line
detection are  the   central X--ray sources in young SNRs which do
not show synchrotron nebulae, like Cas A (Mereghetti, Tiengo \& Israel 2002), 
Puppis A (Zavlin, Tr\"{u}mper \& Pavlov 1999),  and  G 266.2--1.2 (Pavlov et al. 2001b).

\acknowledgments

We thank the EPIC PI Martin Turner for helping us with the scheduling of this EPIC
Guaranteed Time observation. We are also grateful to the VILSPA \xmm team. 
This work was supported by the Agenzia Spaziale Italiana (ASI). ADL acknowledges ASI for a fellowship.


\begin{table}[h]

\caption{{\it  Results of PN fits including absorption lines }}


 \begin{center}

 \begin{tabular}{lccc}

 \hline

\smallskip

          & Blackbody &   Blackbody + Power law  & H atmosphere \\

          & + 2 Gaussians & + 2 Gaussians & + 2 Gaussians   \\

\hline

\smallskip

 & & & \\

N$_H$ (10$^{20}$ cm$^{-2}$) & 3.0$^{+1.5}_{-1.2}$  & 10$^{+4}_{-3}$       &  9$^{+2}_{-1}$   \\
\smallskip
kT     (keV)                &  0.244$\pm$0.005       & 0.219$\pm$0.008    & 0.096$\pm$0.005 \\
\smallskip
R$_{BB}^{(a)}$  (km)        &  1.7$\pm$0.2         &  2.1$\pm$0.3         &  13$\pm$1   \\
\smallskip
$\alpha_{ph} $              &     --               &  2.9$^{+0.4}_{-1.2}$ & --               \\
\smallskip
E$_{1}$  (keV)              &  0.74$\pm$0.02       & 0.73$^{+0.2}_{-0.6}$ &  0.71$^{+0.02}_{-0.04}$ \\
\smallskip
$\sigma_{1}$  (keV)         &  0.13$\pm$0.03       & 0.12$\pm$0.04        &  0.12$\pm$0.04       \\
\smallskip
$EW_{1}$      (eV)          & --90$^{+30}_{-60}$   & --92$^{+30}_{-120}$  & --91$^{+15}_{-60}$    \\
\smallskip
E$_{2}$     (keV)           &  1.36$\pm$0.03       & 1.36$\pm$0.02        &  1.37$\pm$0.02       \\
\smallskip
$\sigma_{2}$  (keV)         &  0.10$\pm$0.05       &  0.08$\pm$0.04       &    0.07$\pm$0.03      \\
\smallskip
$EW_{2}$     (eV)           & --107$^{+36}_{-90}$  & --67$^{+29}_{-54}$   & --63$^{+20}_{-30}$    \\
\smallskip
F$_{0.3-3 keV}^{(b)}$ (erg cm$^{-2}$ s$^{-1}$)  &  1.96$\times10^{-12}$ &  1.96$\times10^{-12}$   &   1.95$\times10^{-12}$     \\
\smallskip
F$_{PL}^{(c)}$ (erg cm$^{-2}$ s$^{-1}$)         &  --                   &  5.5$\times10^{-13}$    &   --    \\
\smallskip
L$_{X}^{(d)}$ (erg s$^{-1}$) &   1.15$\times10^{33}$ &   1.17$\times10^{33}$ & 1.8$\times10^{33}$  \\
\smallskip
$\chi^{2}$/dof              &    1.54            & 1.10               &    1.16  \\
\smallskip
dof                         &   75               & 73                 & 75  \\

\hline
\end{tabular}


\end{center}

$^a$ Radius at infinity for an assumed distance of 2 kpc.

$^b$ Observed flux.

$^c$ Observed flux of the power law component only, 0.3-3 keV.

$^d$ Bolometric luminosity (excluding the power law component) for d=2 kpc.

All the errors are at the 90\% c.l. for a single interesting parameter
\end{table}

\begin{table}[h]

\caption{{\it  Results of phase resolved spectroscopy }}


 \begin{center}

 \begin{tabular}{lcccc }

 \hline

\smallskip
          &    A  &   B &  C &   D \\
          & 0.8--1.   & 0--0.25 & 0.25--0.55  & 0.55-0.8  \\
\hline
\smallskip
 & & & \\
N$_H$ (10$^{20}$ cm$^{-2}$) fixed & 3.0   & 3.0        &  3.0   & 3.0    \\
\smallskip
kT     (keV)    &  0.241$\pm$0.004 &  0.250$\pm$0.004   & 0.245$\pm$0.004  & 0.242$\pm$0.004   \\
\smallskip
R$_{BB}^{(a)}$  (km)   &  1.8        & 1.7    & 1.6    & 1.7    \\
\smallskip
E$_{1}$  (keV) &  0.78$\pm$0.04 &  0.76$\pm$0.04    & 0.74$\pm$0.02 & 0.71$\pm$0.02\\
\smallskip
$\sigma_{1}$  (keV) &  0.12$\pm$0.02   &  0.16$\pm$0.04    & 0.13$\pm$0.02 &  0.08$\pm$0.02  \\
\smallskip
r$_{1}^{(b)}$  &   0.29$\pm$0.08   &   0.28$\pm$0.08   &     0.35$\pm$0.07  &   0.37$\pm$0.11    \\
\smallskip
E$_{2}$ (keV)  & 1.36$\pm$0.04  &  1.37$\pm$0.03   & 1.37$\pm$0.02  &  1.35$\pm$0.02    \\
\smallskip
$\sigma_{2}$  (keV)   &  0.16$^{+0.09}_{-0.04}$   &  0.18$\pm$0.04       &  0.09$\pm$0.02  &  0.06$\pm$0.02      \\
r$_{2}$  &    0.19$\pm$0.09   &  0.30$\pm$0.08    &  0.52$\pm$0.13     &    0.56$\pm$0.16   \\
\smallskip
$\chi^{2}$/dof             &  1.39 &    1.00            & 1.48               &    1.31 \\
\smallskip
dof                         & 63   &   68              & 69                 & 67  \\

\hline

\end{tabular}

\end{center}

 $^a$ Radius at infinity for an assumed distance of 2 kpc.

 $^b$ Relative line depth =  1. -- F (E$_{line}$) / F$_{C}$ (E$_{line}$),
 where F is the total flux and F$_{C}$ is the flux of the continuum only

 All the errors are at the 90\% confidence level for a single interesting parameter

\end{table}

\end{document}